\documentclass[reprint,aps,prl,amsmath,amssymb,floatfix,superscriptaddress]{revtex4-1}

\usepackage{graphicx}
\usepackage{bm}
\usepackage{epstopdf}
\usepackage{siunitx}
\usepackage[pdftex,breaklinks=true,bookmarksopen=true,bookmarksopenlevel=3,bookmarksnumbered=true,colorlinks=true,urlcolor= magenta,citecolor=blue,linkcolor=blue]{hyperref}

\begin{document}

\title{Angular momentum evolution in laser-plasma accelerators}

\author{C. Thaury}
\author{E. Guillaume}
\author{S. Corde}
\author{R.~Lehe}
\author{M. Le Bouteiller}
\author{K. Ta Phuoc}
\affiliation{Laboratoire d'Optique Appliqu\'ee, ENSTA ParisTech - CNRS UMR7639 - \'Ecole Polytechnique, Chemin de la Huni\`ere, 91761 Palaiseau, France}
\author{X. Davoine}
\affiliation{CEA, DAM, DIF, F-91297 Arpajon, France }
\author{J. M. Rax}
\author{A. Rousse}
\author{V. Malka}

\affiliation{Laboratoire d'Optique Appliqu\'ee, ENSTA ParisTech - CNRS UMR7639 - \'Ecole Polytechnique, Chemin de la Huni\`ere, 91761 Palaiseau, France}

\begin{abstract}
The transverse properties of an electron beam are characterized by two quantities, the emittance which  indicates the electron beam extent in the phase space and the angular momentum which allows for non-planar electron trajectories.
Whereas the emittance of electron beams produced in laser-plasma accelerator has been measured in several experiments, their angular momentum has been scarcely studied.
It was demonstrated that electrons in laser-plasma accelerator carry some angular momentum, but its origin was not established. Here we identify one source of angular momentum growth and we present experimental results showing that the angular momentum content evolves during the acceleration.
\end{abstract}

\maketitle 

Since the first observation of quasi-monoenergetic electron beams in 2004~\cite{Nature2004Faure,Nature2004Geddes,Nature2004Mangles}, the features of laser-plasma accelerator in the bubble/blow-out regimes~\cite{APB2002Pukhov,PRL2006Lu} have been extensively studied, and constantly improved.  High quality electron beams can now be accelerated up to the giga-electronvolt level~\cite{NatPhys2006Leemans}. In the few hundred of mega-electronvolt (MeV) range,  electron beam with 1\% energy spread~\cite{PRL2009Rechatin1} and few kiloamperes peak current ~\cite{NatPhys2011Lundh} can be reliably produced. The electron source size is  a fraction of micrometer~\cite{corde_prl_2011a}, the electron divergence is a few mrad, and the normalized emittance is of the order of $1\pi$ mm.mrad or smaller~\cite{PRL2010Brunetti,2012PRL_Plateau,weingartner2012}.
Yet, in ten years of intensive investigations, one fundamental property of laser-plasma accelerated electron beams, the beam angular momentum, has  been scarcely studied.

A couple of  experiments showed that  laser-plasma accelerated beams can carry some angular momentum~\cite{PRL2006TaPhuoc}, but  little effort has been made to elucidate its origin. Injection models predict that electrons should be injected in the accelerator with a zero angular momentum~\cite{PRL2009Kostyukov}. The fact that electrons can have a significant angular momentum thus means that either it grows during the acceleration, or injection models are incomplete.
In this letter, we show that a non-perfectly symmetric laser pulse can create an asymmetric plasma cavity that, in turn, induces an evolution of the electrons' angular momentum during the acceleration. This explanation for the origin of the angular momentum is supported by experimental results and simulations.

Laser pulses with aberrated wavefronts are known to drive anisotropic plasma cavities~\cite{glinec2008,mangles2009apl,PRL2010POPP}. This anisotropy changes the electron trajectories in the plasma, and hence modifies the properties of the  accelerated electrons and  the X-rays they emit.
In our experiment, the focal spot of the laser is observed to be elliptical (mainly because of astigmatism),  with  a typical  eccentricity $\approx 0.6$, we can therefore consider that the plasma cavity is also elliptical. Assuming that the ellipse axes are along the $\bm{x}$ and $\bm{y}$ axes,
the transverse forces in such a cavity are 
\begin{align}
F_x&=-\alpha m (1+\epsilon/2) \omega_p^2 x/2 \label{eq:fx}\\ 
F_y&=-\alpha m (1-\epsilon/2) \omega_p^2 y/2 \label{eq:fy}
\end{align} with $m$ the electron mass, $\omega_p$ the plasma frequency, $\alpha\leq 1$ a coefficient describing a possible deviation from the nominal transverse force of a fully evacuated ion cavity and $\epsilon <1$ a coefficient which quantifies the asymmetry of the transverse force.

\begin{figure}
	\centering
		\includegraphics{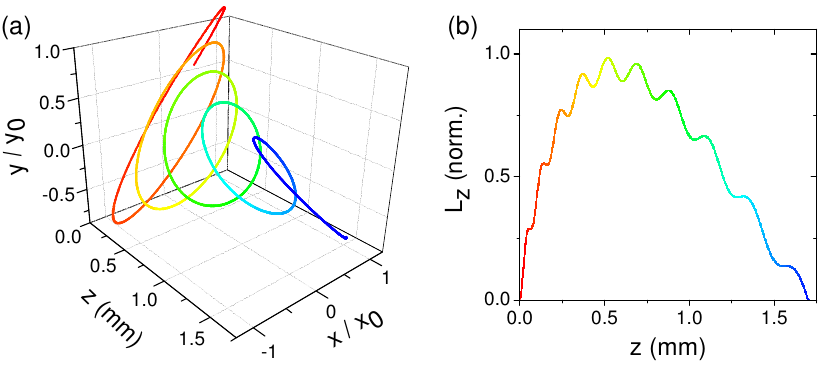}
	\caption{Evolution of the angular momentum in an elliptical cavity. (a) Trajectory of an electron. (b) Angular momentum as a function of the longitudinal position $z$  (the color is a function of $z$). The parameters are $\epsilon=0.2$, $\gamma_i=25$, $\kappa=10^{14}$ s$^{-1}$, $n_e=8\times10^{18}$ cm$^{-3}$ and $\alpha=0.5$),}
	\label{fig:fig1}
\end{figure}

 For adiabatic acceleration, the equation of motion can be integrated, leading to 
\begin{align}
x&= x_0(\gamma_i/\gamma)^{1/4} \sin [(1+\epsilon/2)^{1/2}\phi(t)+\phi_{x0}]\\
y&= y_0(\gamma_i/\gamma)^{1/4} \sin [(1-\epsilon/2)^{1/2}\phi(t)+\phi_{y0}]
\end{align}
 with $\gamma$ the electron Lorentz factor, $\phi(t)=\int_0^t [\alpha/2\gamma(t')]^{1/2}\omega_p dt'$,  $(x_0, y_0)$  the initial position, ($\phi_{x0},\phi_{y0}$) the initial phases, and $\gamma_i $ the value of $\gamma$ at injection. For $\epsilon \neq 0$, electrons in the cavity oscillate with different frequencies along $\bm{x}$ and $\bm{y}$.  As a result,  their trajectories which are initially planar for electrons with zero angular momentum ($\phi_{x0}=\phi_{y0}=\phi_0$)\footnote{For linear polarization, electrons are likely injected without angular momentum,  while for circular polarization, electrons can be injected with a weak $L_z$. The evolution of $L_z$ is then similar in both cases.}, progressively become helicoidal, before reverting to planar trajectories  every time  $\phi(t)=k\pi/(1-\sqrt{1-\epsilon})$ with $k$ an integer. In other words the angular momentum $L_z=x p_y-y p_x$ changes in time. For $\epsilon <<1$   it can be written as 
\begin{equation}
\frac{L_z}{m\omega_p }=x_0 y_0 \sqrt{\frac{\alpha\gamma_i}{2}}\left( \sin \frac{\epsilon}{2}\phi-\frac{\epsilon}{4}\sin\left[2\left(\phi +\phi_0\right)\right]\right)\text{.}
\label{eq:eq1}
\end{equation}
The first term accounts for slow variations of $L_z$; it is responsible for the transition from planar to helicoidal trajectories. The second term corresponds to high frequency oscillations. It is of low amplitude and  can be neglected in a first approximation (see Fig.~\ref{fig:fig1}). 
Equation~(\ref{eq:eq1}) also shows  that, for a given initial radius $r_0=(x_0^2+y_0^2)^{1/2}$, $L_z$ is maximum when $x_0=y_0$, that is when electrons are initially in a plane at $45^\circ$ to the ellipse axes, whereas $L_z$ remains 0 at all times for on-axis electrons. Note that the losses of energy and  angular momentum by radiation are negligible and were not considered in the derivation of Eqs. (3-5).  The angular momentum conservation is ensured by  the fact  that  electrons and ions from the cavity sheath acquire some angular momentum~\cite{lopez1975}.

\begin{figure}
	\centering
		\includegraphics{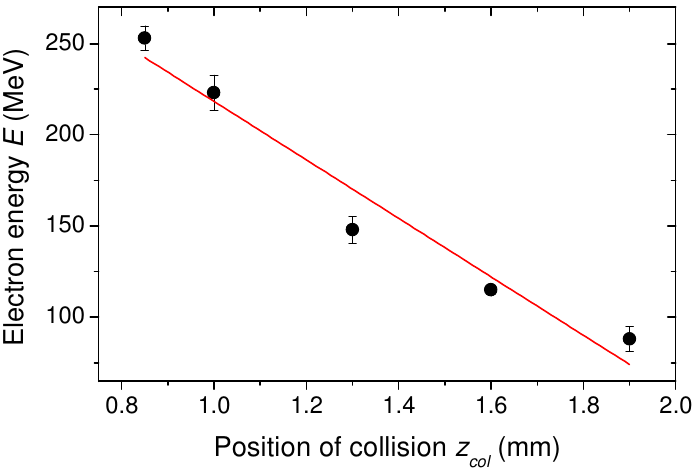}
	\caption{Electron beam energy as a function of the collision position. The position $z_{col}=1.5$ mm corresponds to a collision occurring at the center of the gas jet. The line is a linear fit and the error bar corresponds to the standard error.}
		\label{fig:fig2}
\end{figure}

For a given $(x_0,y_0)$, the angular momentum reaches its peak value when $\phi(t)=\pi/\epsilon$. Assuming for simplicity $\gamma=\gamma_i+\kappa t$ (which is consistent with Fig.~\ref{fig:fig2}), we find for $\epsilon=0.2$ and typical laser-plasma parameters ($\gamma_i=25$, $\kappa=10^{14}$ s$^{-1}$, $n_e=8\times10^{18}$ cm$^{-3}$ and $\alpha=0.5$), that $L_z$ is maximum after an acceleration of $\approx 600$ $\mu$m (this length is reduced when $\epsilon$ is increased).  As effective acceleration lengths in experiments are generally about or larger than 1 mm, the slow oscillations of $L_z$ should be observable, assuming that the acceleration length can be precisely controlled. In our experiment, we achieved this control through the colliding pulse injection scheme~\cite{Nature2006Faure}. Information on the angular momentum content of the beam is then obtained from the analysis of betatron X-rays  emitted during the transverse oscillations of the accelerated electrons~\cite{PRL2004Rousse,RMP_corde}.

 The experiment was performed at Laboratoire d'Optique Appliqu\'ee with the `Salle Jaune' Ti:Sa laser system. Two synchronized 35 fs FWHM laser pulses were used: the pump pulse that  drives the accelerating plasma wave contained 900 mJ and the injection pulse that triggers the injection into the main pump pulse wakefield contained 100 mJ . The two pulses had the same linear polarization. They were focused onto a 3 mm supersonic helium gas jet where they collided at a 135 degrees angle. The pump pulse (respectively the injection pulse) had a mean FWHM focal spot size of  18 $\mu$m (respectively 22 $\mu$m) and a normalized vector potential amplitude of $a_0=1.3$ (respectively $a_0=0.4$). Electron spectra and x-ray angular profiles were measured simultaneously in a single shot. The electron spectrometer consisted of a permanent bending magnet (1.1 T over 10 cm) combined with a phosphor screen imaged on a 16 bits CCD camera. X-ray profiles were obtained from an x-ray CCD placed on axis at 90 cm from the gas jet, behind a 25 $\mu$m Be filter. In this experiment, the electron plasma density was $n_e=8\times10^{18}$ cm$^{-3}$, which corresponds, for our experimental parameters, to an interaction regime where electrons are not self-injected in the wakefield. Consequently, electrons and x-rays were observed only when both laser pulses overlapped in time and space.

\begin{figure*}
	\centering
		\includegraphics[]{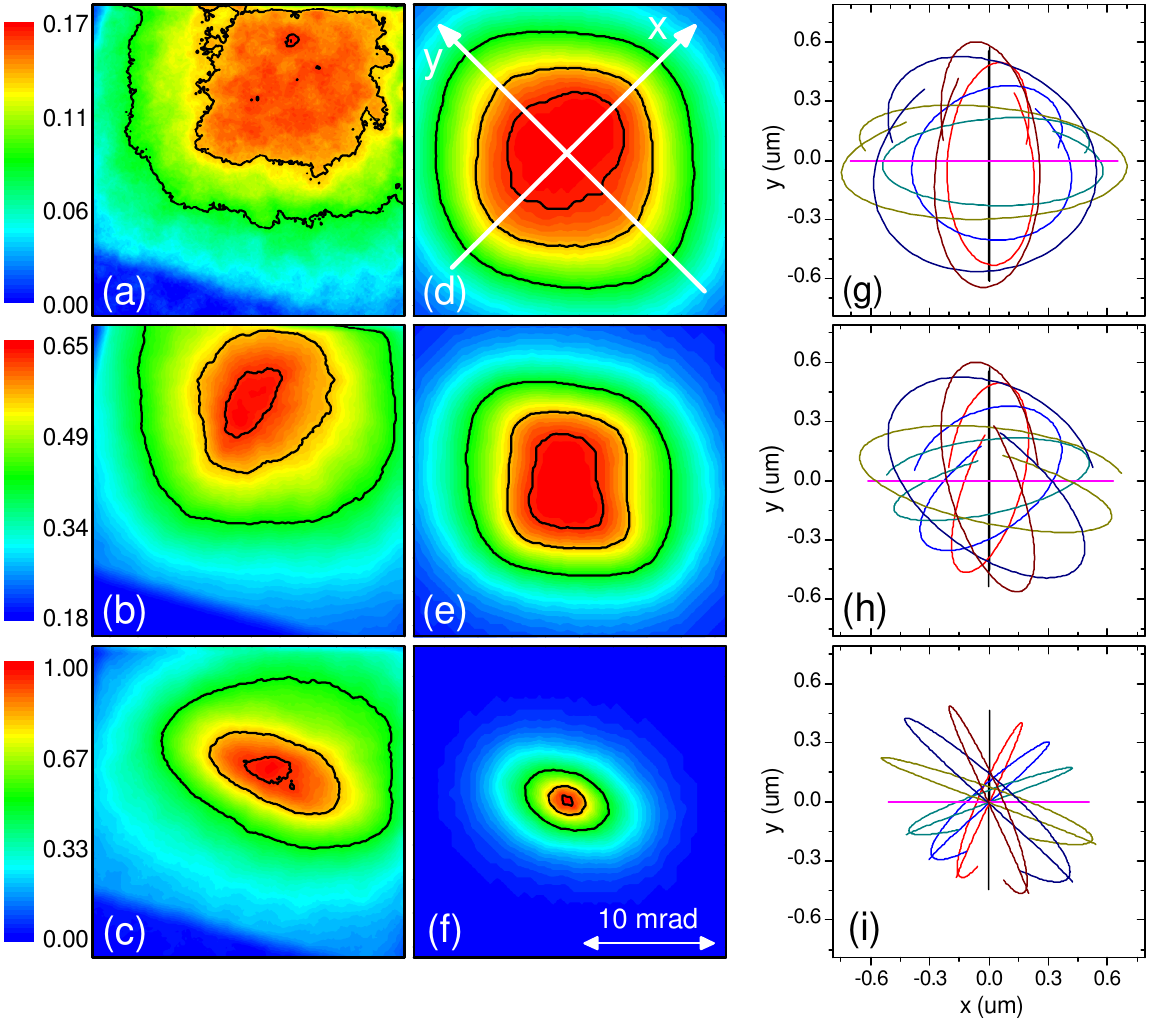}
		\caption{X-ray angular profiles for different acceleration lengths, from the experiment (a,b,c) or from a test-particle simulation (d,e,f) and typical  trajectories of 8 electrons (g,h,i). The electron energy is 120 MeV in (a,d,g), 160 MeV in (b,e,h) and 260 MeV in (c,f,i). A 0.9 mrad mean filter was applied on the experimental images. Contour lines at 50\%, 80\% and 95\% of the peak intensity are superimposed on the angular profiles. Only the last betatron period is plotted in (g-i).  In the simulation,  $\epsilon =0.2$,  $\alpha=0.55$, and all electrons have a zero initial angular momentum. The beam is initially matched with a phase $\phi_{0}$ uniformly distributed between 0 and $2\pi$. It has a flat elliptical transverse distribution oriented at 36$^\circ$ to the x-axis, with an aspect ratio of 0.5. }
			\label{fig:fig3}
\end{figure*}

 Experimental results in Fig.~\ref{fig:fig2} demonstrate that the electron energy $E$ can be tuned from $\approx 90$ MeV to $\approx 250$ MeV, by adjusting the collision position and hence the acceleration length.  The linear fit shows that $\gamma$ can be reasonably approximated by a linear function, leading to $E\approx -160z_{col}[mm] +\text{const.}$ and $\kappa \approx 10^{14}s^{-1}$.
 To study the evolution of the angular momentum content of the beam, we now focus on X-ray measurements. Figure~\ref{fig:fig3}a-c shows single-shot X-ray angular profiles corresponding to electron energies of 120 MeV (a), 160 MeV (b) and 260 MeV (c). The X-ray divergence  decreases when $E$ increases, due to a reduction of  the electron beam divergence in the acceleration ($\theta_{electrons}\propto \gamma^{-3/4}$). More interestingly, we also observe that the X-ray profiles evolve from somewhat rectangular and flat shapes to elliptical shapes.

\begin{figure}
\begin{center}
\includegraphics{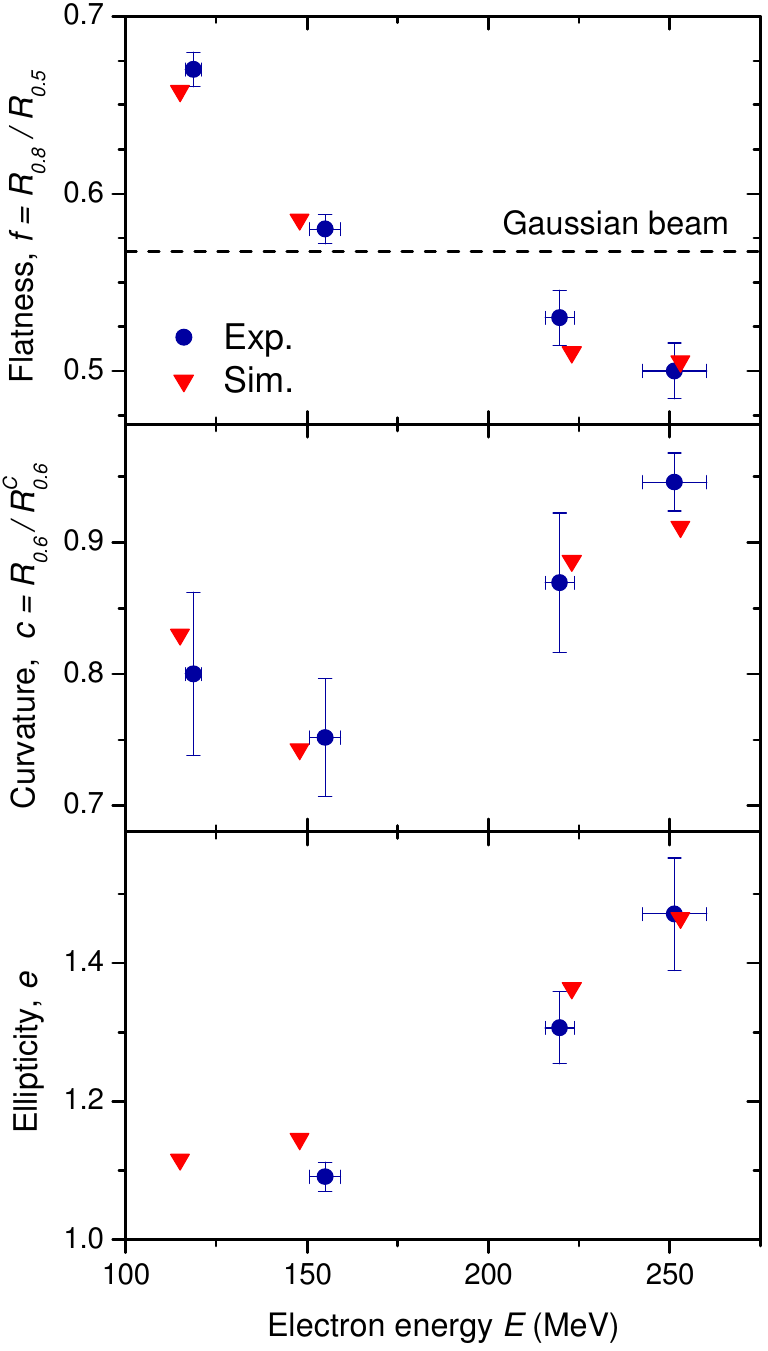}
\caption{Variation of the flatness, the curvature and the ellipticity with the electron energy.  The dots corresponds to experimental data and the triangles to simulations. Each experimental point represents an average over more than 8 shots obtained for the same acceleration length. Error bars indicate the standard error of the mean. For $E=120$ MeV the ellipticity cannot be computed accurately because X-ray profiles are cut  (see e.g. Fig.~\ref{fig:fig3}a).  The simulation parameters are the same as in Fig.~\ref{fig:fig3}.}
\label{fig:fig4}
\end{center}
\end{figure}

To quantify this evolution, we define 3 variables, the ellipticity $e$,  the flatness $f$ and the curvature $c$. The ellipticity of the X-ray profile is calculated from an ellipse fit of the 50\% contour line. It is defined as the ratio of the major to the minor radii. The flatness $f$ is given by the ratio of $R_{0.8}$ to $R_{0.5}$, with $R_{0.8}$ and $R_{0.5}$ the mean radii at 80\% and 50\% of the peak intensity ($f=1$ for a top-hat beam and $f=0.57$ for a Gaussian beam). Lastly,  $c=R_{0.6}/R_{0.6}^C$ with $R_{0.6}$ the mean radius at 60\% and $R_{60\%}^C$ the mean curvature radius of the 60\% contour line (computed using the algorithm described in~\cite{nguyen}).  It follows that  $c=1$ for a perfect circle, while $c=0$ for a square. Figure~\ref{fig:fig4} shows that $c$ and $e$ increase with $E$, while $f$ decreases when $E$ increases. This confirms the trend observed in Fig.~\ref{fig:fig3}a-c.

This behavior can be explained by an evolving angular momentum. In the wiggler approximation ($k_pr_0(\gamma/2)^{1/2}\gg1$ with $k_p=\omega_p/c$ ), electrons with $L_z=0$ radiate an elliptical X-ray beam of divergences $\theta_\parallel=k_p r_0/\sqrt{2\gamma}$  along the oscillation direction, and $\theta_\perp=1/\gamma$ in the direction orthogonal to the oscillations~\cite{RMP_corde}. The measured X-ray profile is an incoherent sum of the contributions of all electrons. For an isotropic electron distribution with $L_z=0$,  the sum  results in a circular profile consisting of  a  central peak with a divergence of $\theta_\perp$ surrounded by a halo with a divergence of $\theta_\parallel$. Anisotropic electron distributions with a preferential oscillation direction lead to elliptical profiles with a central peak. These features are consistent with X-ray beams obtained for the longer acceleration length ($E=250$ MeV). In contrast, electrons with a maximal $L_z$ have  circular orbits and emit annular X-ray beams of angular radius $k_pr_0/\sqrt{2\gamma}$ and  thickness $1/\gamma$~\cite{PRL2006TaPhuoc} . Summing over electrons of different $r_0$ results in a flatter beam than for $L_z=0$ with no central peak. For electrons satisfying Eq.~(\ref{eq:eq1}) with $\epsilon\phi=\pi$ and an initial amplitude $r_0$, the orbits are planar with an oscillation amplitude $r_0$ for electrons initially located on the $\bm{x}$ and $\bm{y}$ axes , and circular  with a radius $r_0/\sqrt{2}$ for electrons such as $x_0=y_0$.  More precisely, all trajectories are contained in a square of side length $\sqrt{2}r_0$ with diagonals along the axes, as shown in Fig.~\ref{fig:fig3}g.  As a result, the X-ray beam obtained by summing over the contribution of electrons of different initial positions and velocities has a square shape and a relatively flat profile, similar to X-ray beams measured for the shorter acceleration length ($E=120$ MeV). This simple analysis thus suggests that the angular momentum, in our experiment, is maximum for $E\approx120$ MeV and that it decreases as the electron energy increases further.

Assuming that angular momentum variations are due to the cavity ellipticity and that the electron initial transverse distribution does not depend on the injection position, the results indicate that the phase difference between the oscillations along the $\bm{x}$ and $\bm{y}$ axis reaches $\pi/2$ for $E\approx 120 MeV$, that is for an acceleration length of $600$ $\mu$m.   According to Eq.~(\ref{eq:eq1}), this implies that the cavity ellipticity is  $\epsilon \approx 0.15 \alpha^{-1/2}$. As electrons are further accelerated, the difference of phase keeps increasing,  which explains why the angular momentum tends to decrease in the experiment. 

The  X-ray beam ellipticity  at high energy indicates that the electron distribution is anisotropic~\footnote{The ellipticity cannot be due to the interaction of the accelerated  electrons with the laser,  because the ellipse  major axis is not parallel to the laser polarization}. If electrons were mostly distributed along the $\bm {x}$ or $\bm{y}$ axis, they would never develop a significant angular momentum and only elliptical beams would be obtained. In contrast, the fact that X-ray profiles with a square shape are measured indicates that electrons are preferentially injected with amplitudes $x_0= y_0$, because only electrons with a maximum $L_z$ can lead to non-elliptical emissions. This implies that the ellipse axes at high electron energy should make a 45$^\circ$ angle with  the diagonals of the square profiles observed at low energy (the diagonals should be along the  $\bm {x}$ and $\bm{y}$ axis). Accordingly, we experimentally measured, for $E\approx120$ MeV, a mean angle $\Psi_s=45(+90)\pm1^\circ$ between the square diagonals  and the horizontal axis, and, for $E\approx250$ MeV a mean angle $\Psi_e=1(+90)\pm9^\circ$ between the ellipse axes and the horizontal axis (this indicates that in the experiment the cavity axes  are at $\pm45^\circ$ from the horizontal axis) . The angle $\Psi_s$ was observed to be very stable shot-to-shot, while  $\Psi_e$  drifted from $-13(+90)^\circ$ up to $28(+90)^\circ$. The ellipse axes were also observed to swap in time. The reason is that  $\Psi_s$ is determined by  the orientation of the elliptical cavity, which should not change significantly shot-to-shot, while $\Psi_e$ depends on the distribution in the transverse phase space of injected electrons, which is more sensitive to laser fluctuations. Small changes in the injected distribution can, for instance, originate from ionization induced refraction of the injection pulse~\cite{rae1993optcom,chessa1999prl}, and from the stochastic nature of the heating process~\cite{sheng2002prl}. 

To confirm this analysis, we performed test-particle simulations, using
 the experimental energy spread, the longitudinal acceleration force measured in Fig.~\ref{fig:fig2} , the transverse force from Eqs.~(\ref{eq:fx},\ref{eq:fy}) and the source size calculated from  the X-ray spectrum~\cite{corde_prl_2011a}. We scanned a large range of $\epsilon$ and of initial electron distributions, and we found that an agreement with experiment data is obtained only for $\epsilon \approx 0.2$ and $\alpha\approx 0.55$,   a matched electron beam and an elliptical  initial transverse distribution, oriented close to the  $x=y$ line.  Typical electron trajectories obtained in this case are plotted in Fig~\ref{fig:fig3}g-i; as expected, they evolve from helicoidal to planar trajectories as the electron energy increases. Figures~\ref{fig:fig3}d-f and~\ref{fig:fig4} show that the X-ray angular profiles, calculated using the general formula for the radiation emitted by  relativistic
electrons~\cite{jackson},  reproduce accurately the experimental divergence as well as the behavior of $f$, $c$ and $e$, except for the divergence at high electron energy.  This discrepancy could be due to the interaction of the electron beam with the laser pulse at the end of the acceleration. Another limitation of the model is the assumption of a steady and perfectly elliptical cavity. In particular, because of different self-focusing dynamics in the two transverse directions, the ellipticity can vary in time. This can modify the extrema of $L_z$, as well as the electron energy at which these extrema are obtained

Apart from the difference of divergence, the good  overall agreement between the experiment and test-particle simulations indicates that our simple model includes most of the relevant physics. Simulations can therefore be used to estimate the transverse emittance. This leads in our case to a normalized emittance of about $1\pi$ mm.mrad.  Since the simulation fits  both the measured X-ray spectra and the angular profiles, they provide  both $\alpha$ and the emittance. Estimating these two quantities is essential in  betatron based emittance measurements  because the inferred emittance varies as $\alpha^{3/2}$. Assuming that the cavity is fully evacuated ($\alpha=1$), as done for instance in~\cite{2012PRL_Plateau}, may thus  lead to an underestimate of the emittance.


In conclusion, we demonstrated that the angular momentum content of an electron beam, accelerated in an anisotropic cavity, is time varying, and we provided experimental evidence of such variations. These results have important consequences for several emittance measurement techniques. 
Neglecting the angular momentum can in particular result in unreliable estimates of the emittance in X-ray based measurements, because the spatial properties of the X-rays strongly depend on the angular momentum content. In addition, the angular momentum growth is a source of fluctuations, which can, for instance, induce large shot-to-shot  changes in betatron profiles. To avoid such effects, the laser pulse should be free of aberration in order to produce an axisymmetric cavity. Conversely, it could be beneficial to take advantage of an asymmetric cavity to manipulate the shape of the X-ray beams and produce a radiation with a net angular momentum. This would require a precise control of the transverse distribution of injected electrons.

{\textbf{Acknowledgments}}
\noindent
This work was supported by the European Research Council (PARIS ERC, Contract No. 226424), EuCARD/ANAC, EC FP7 (Contract No. 227579), EuCARD2/ANAC2 EC FP7 (Contract No. 312453) and Agence National pour la Recherche (FENICS ANR-12-JS04-0004-01).
%

\end{document}